\newcommand{\op}[1]{\hat{#1}}
\newcommand{\openone}{\leavevmode\hbox{\small1\normalsize\kern-.33em1}}

\documentclass[10pt]{iopart}
\usepackage{iopams,bm,cite,graphicx,harvard,url,times}

\begin{document}

\title{Depolarization for quantum channels with higher symmetries}

\author{A B Klimov$^1$ and L L S\'{a}nchez-Soto$^2$} 

\address{$^1$ Departamento de F\'{\i}sica,
Universidad de Guadalajara, 
44420~Guadalajara, Jalisco, Mexico}

\address{$^2$ Departamento de \'Optica,
Facultad de F\'{\i}sica, Universidad Complutense, 
28040~Madrid, Spain}

\date{\today}

\begin{abstract}
  The depolarization channel is usually modelled as a quantum
  operation that destroys all input information, replacing it by a
  completely chaotic state. For qubits this has a quite intuitive
  interpretation as a shrinking of the Bloch sphere. We propose a way
  to deal with depolarizing dynamics (in the Markov approximation) for
  systems with arbitrary symmetries.
\end{abstract}

\bigskip

Quantum systems evolve via unitary operations determined by the
Schr\"{o}dinger equation. This is also true for composite systems,
e.g., a small quantum system $\mathcal{S}$ that is surrounded by 
some environment $\mathcal{E}$, with which it interacts.  However, 
while  the evolution of the total system is described by a unitary 
operation $\op{U}_{\mathcal{SE}} (t)$, the dynamics of the system 
$\mathcal{S}$ alone (obtained by tracing out the uncontrollable 
degrees of freedom of $\mathcal{E}$) is, in general, no longer
unitary~\cite{Breuer:2007,Weiss:2008,Alicki:2007}. The
system-environment interaction leads to entanglement between 
them, and this is reflected into the fact that $\op{U}_{\mathcal{SE}} (t)  
\neq \op{U}_{\mathcal{S}} (t) \otimes \op{U}_{\mathcal{E}} (t)$. From 
the perspective of quantum information processing, such an 
interaction  is undesirable and causes errors and noise in the 
system~\cite{Nielsen:2000}.

The dynamics of $\mathcal{S}$ can be described by a finite-time
trace-preserving completely positive map
(CPM)~\cite{Stinespring:1955,Choi:1972} that transforms input states
$\op{\varrho}_{\mathrm{in}} = \op{\varrho} (0)$ into output states
$\op{\varrho}_{\mathrm{out}} = \op{\varrho} (t)$, i.e.,
\begin{equation}
  \varrho_{\mathrm{in}} \mapsto \op{\varrho}_{\mathrm{out}} =  
  \mathcal{E}_{t} (\op{\varrho}_{\mathrm{in}}) \, ,
\end{equation} 
which is also known as a quantum channel. One may think of the
environment as extracting information from the system, as it will
typically map pure states into mixed states.

This noise process can be also described by a quantum operation
involving only operators of the system of interest. This is called a
Kraus decomposition~\cite{Kraus:1983} and has the form (omitting all
the unnecessary subscripts)
\begin{equation}
  \label{eq:Krausdec}
  \mathcal{E} ( \op{\varrho} ) = 
  \sum_{r} \op{K}_{r} \, \op{\varrho} \, \op{K}_{r}^\dagger \, ,
\end{equation}
where the Kraus operators satisfy the condition
\begin{equation}
  \sum_{r} \op{K}_{r} \, \op{K}_{r}^\dagger = \op{\openone} \, ,
\end{equation}
which ensures that unit trace is preserved for all times.
 
A channel is Markovian when the coupling coupling of the system
$\mathcal{S}$ with the environment $\mathcal{E}$ can be treated under
the Markov and Born approximations. The evolution of the state at a
given instant is then fully determined by the state at that instant,
so the process ha no ``memory'' of its past. This is a commonly used
approximation in quantum optics and leads to the well-known Lindblad
form of a master equation~\cite{Lindblad:1976,Gorini:1976}.  For these
Markovian channels, one can always write
\begin{equation}
  \mathcal{E}_{t} (\op{\varrho}) = e^{\mathcal{L} t} \,  
  \op{\varrho} (0) \, ,
\end{equation}
where the Lindblad superoperator $\mathcal{L}$ is
\begin{equation}
  \label{eq:Lindblad}
  \mathcal{L} (\op{\varrho} )  = - i [\op{H}, \op{\varrho} ] +
  \frac{1}{2} \sum_{r} a_{r} \left ( 
    [ \op{L}_{r},  \op{\varrho} \op{L}_{r}^\dagger ] + 
    [ \op{L}_{r} \op{\varrho} , \op{L}_{r}^\dagger ] 
  \right ) \, .
\end{equation}
Here $\op{H}$ is the Hamiltonian of the undamped system $\mathcal{S}$,
$\op{L}_{r}$ are system operators defined to model the effective
dissipative interaction with the environment, and $a_{r} \ge 0$
are constants that account for decoherence rates. In this form,
$\mathcal{L}$ appears as the generator of a CMP. In fact, expanding
equation~(\ref{eq:Lindblad}) to first order in the short-time interval
$\tau$ one can immediately find the corresponding Kraus
operators~\cite{Shabani:2005}
\begin{eqnarray}
  \label{eq:Lininf}
  \op{K}_0 & = &  \openone  - \tau 
  \left ( i \op{H} + 
    \frac{1}{2} \sum_{r}  \op{L}_{r}^\dagger \, \op{L}_{r} 
  \right )  \,  \nonumber \\
  & & \\
  \op{K}_{r} & = & \sqrt{\tau} \,  \op{L}_{r} \, . \nonumber
\end{eqnarray}

For obvious reasons, the characterization and classification of 
these maps have attracted a lot of interest in recent
years~\cite{Keyl:2002}.  The majority of the results obtained 
so far relate to two specific classes: the qubit channels and 
the bosonic Gaussian channels~\cite{Caruso:2007}. In this paper,
we shall be mainly concerned with the former.

In classical computation, the only error that can occur is the bit
flip $0 \leftrightarrow 1$. In quantum computation, however, the
existence of superposition states brings also the possibility of other
basic errors for a single qubit.  They are the phase flip and the
bit-phase flip. The first changes the phase of the state, and the
latter combines phase and bit flips.  The set of Kraus operators for
each one of these channels is given by
\begin{equation}
  \label{eq:Ksqb}
  \op{K}_{0} = \sqrt{1 - p/2} \, \op{\openone} \, ,
  \qquad
  \op{K}^{j}_{1} = \sqrt{p/2} \, \op{\sigma}_{j} \, ,
\end{equation}
where $j = x$ gives the bit flip, $j = z$ the phase flip and $j = y$
the phase-bit flip. They can also interpreted as corresponding to a
probability $1 - p/2$ of remaining in the same state, and a
probability $p/2$ of having an error. It is not difficult to
represent these channels in terms of a master equation of the
form (\ref{eq:Lindblad}): the associated Lindblad superoperator
$\op{L}^{j}_{1}$ turns out to be the corresponding Pauli matrix
$\op{\sigma}_{j}$.

The environment-induced noise is modelled using various simplified
approaches. Here, we concentrate in the depolarization channel, which
employs unbiased noise generating bit flip errors and phase flip
errors and is represented by the CPM~\cite{Bennett:1997}
\begin{equation}
  \label{eq:depolqbit}
  \mathcal{E} (\op{\varrho} ) = 
  ( 1 - \op{\varrho} ) + \frac{p}{3} 
  (\op{\sigma}_{x} \, \op{\varrho} \, \op{\sigma}_{x} 
  + \op{\sigma}_{y} \, \op{\varrho} \, \op{\sigma}_{y} 
  + \op{\sigma}_{z} \, \op{\varrho} \, \op{\sigma}_{z} ) \, .
\end{equation}
Note, however, that, since the three interaction channels
corresponding to bit error, flip error, and phase error do not
commute, one could rightly argue that the incoherent addition of these
channels in the CMP~(\ref{eq:depolqbit}) is, at least, questionable.

The Kraus decomposition of this channel reads as
\begin{eqnarray}
  \label{eq:Krausdepol}
  & \op{K}_{0} = \sqrt{1-p} \, \op{\openone} , & \nonumber \\
  & & \\
  & \displaystyle
  \op{K}_{1} = \frac{p}{3} \, \op{\sigma}_{x} , \qquad
  \op{K}_{2} = \frac{p}{3} \, \op{\sigma}_{y} ,    \qquad
  \op{K}_{3} = \frac{p}{3} \, \op{\sigma}_{z} \, , \nonumber
\end{eqnarray}
while the associated Lindblad equation is
\begin{equation}
  \label{eq:depchlme}
  \dot{\op{\varrho}} =  - \Gamma \left (
    \op{\varrho} - \frac{1}{2} \op{\openone} \right ) \, ,  
\end{equation}
where we have omitted the free evolution of the system, since it is
irrelevant for our purposes here, and $\Gamma$ is a constant. If we
use the standard Bloch parametrization for the density matrix
\begin{equation}
  \label{eq:Bloch}
  \op{\varrho} = \frac{1}{2} \left ( \op{\openone} + 
    \bm{s} \cdot \op{\bm{\sigma}} \right ) \, ,
\end{equation}
we immediately get
\begin{equation}
  \label{eq:predam}
  \dot{\bm{s}} =  -  \Gamma \bm{s} \, , 
\end{equation}
which clearly shows that the action of the channel is to contract the
sphere with a lifetime $\Gamma^{-1}$. This is precisely the idea
behind depolarization: for long times the system will end in a fully
depolarized or chaotic state, whose density matrix is diagonal
$\op{\varrho}_{\mathrm{unpol}} = \frac{1}{2} \op{\openone}$.

For $n$ qubits, a possible extension is to assume that the error
operators can be represented by the $n$-qubit Pauli
group~\cite{Nielsen:2000}
\begin{equation}
  \label{eq:Pn}
  \mathcal{P}_n = \{ \op{\openone}, \op{\sigma}_{x},  
  \op{\sigma}_{y}, \op{\sigma}_{z} \}^{\otimes n} \, ,
\end{equation}
where $\otimes n$ denotes the $n$-fold tensor product. This means that
each qubit is acted by the identical independent depolarizing
channels.  However, alternative models, such as collective or
correlated depolarization, have been
proposed~\cite{Reina:2002,Banaszek:2004,Ball:2004,Ball:2005,Ban:2006},
which do not fit in this simple extension. The point we wan to stress
is that the final result depends obviously of the symmetry of the
channel and this, in turn, depends on physical considerations.

The purpose of this work is to address this problem and identify the
proper depolarizing channel for more general situations. To this end,
consider a system whose Hamiltonian is a function of a symmetry Lie
algebra $\mathfrak{A}$.  In order to maintain the discussion as simple
as possible, we assume that this algebra is semisimple and denote by
$\Delta$ the set of nonzero roots~\cite{Erdmann:2006dg}.  We use the
standard Cartan-Weyl basis $\{ \op{h}_{i}, \op{e}_{\alpha}\}$ in terms
of which we have the commutation relations
\begin{equation}
  \label{eq:CR} 
  [ \op{h}_{i}, \op{h}_{j}] = 0 \, , 
  \qquad
  [ \op{h}_{i}, \op{e}_{\alpha}] = \alpha (\op{h}_{i}) \, 
  \op{e}_{\alpha} \, , 
  \qquad
  [\op{e}_{\alpha} , \op{e}_{\beta}] = N_{\alpha \beta} \, 
  \op{e}_{\alpha + \beta} \, ,
\end{equation}
where the last one is valid only when $\alpha + \beta \in \Delta$. The
operators $\{ \op{h}_i \}$ ($i$ runs from 1 to $\ell$, where $\ell$ is
the rank of the group) constitute the Cartan subalgebra and may be
taken diagonal in any irreducible representation. On the other hand,
$\{ \op{e}_{\alpha}, \op{e}_{-\alpha} \}$ are raising and lowering
operators and we can always choose $\op{e}_{\alpha}^{\dagger} =
\op{e}_{- \alpha}$.
 
The Hilbert space decomposes also into finite-dimensional subspaces
\begin{equation}
  \label{eq:7}
  \mathcal{H}= \bigoplus_{\lambda}\mathcal{H}_{\lambda}
\end{equation}
so that in each $\mathcal{H}_{\lambda}$ the operators of $\mathfrak{A}$ 
act irreducibly. Let $|\mathbf{h}; \lambda \rangle $ be an orthonormal
basis in $\mathcal{H}_{\lambda}$, where $\mathbf{h}= (h_{1}, \ldots,
h_{j}, \ldots, h_{\ell})$. Then, we have that
\begin{equation}
  \op{h}_{j} |\mathbf{h}; \lambda \rangle = 
  h_{j} | \mathbf{h}; \lambda \rangle  \, .
\end{equation}
Notice also that the density operator of any quantum unpolarized
state can be written in terms of these invariant subspaces
\begin{equation}
\label{denunpol}
\op{\varrho}_{\mathrm{unpol}} = \bigoplus_{\lambda}
r_{\lambda} \,  \op{\openone}_{\lambda} \, ,
\end{equation}
where $\op{\openone}_{\lambda}$ denotes the unity in the corresponding
subspace and all the coefficients $r_{\lambda}$ are real and nonnegative
and they fulfill the unit-trace condition.

Since all the error operators are in the algebra $\mathfrak{A}$, 
the only possible Kraus operators are also elements of $\mathfrak{A}$, 
and they induce a local Lindblad equation fully analogous
to~(\ref{eq:Lindblad}). Very different processes can be described 
in terms of this master equation. The first one is what can be 
called a pure dephasing channel, represented by
\begin{equation}
  \label{eq:deph}
  \op{L}_{r} = \op{h}_{r} \, .
\end{equation}
Such a map obviously preserves the diagonal operators: $\mathcal{L} (
\op{h}_{r} ) = 0$, and the asymptotic limit of the channel is given by
\begin{equation}
  \lim_{n \rightarrow \infty} \mathcal{E}^{n} ( \op{\varrho} ) = 
  \sum_{\mathbf{h},\lambda} \varrho_{\mathbf{h} \mathbf{h}} (\lambda ) \, 
  |\mathbf{h};\lambda \rangle 
  \langle \mathbf{h}; \lambda | \, , 
\end{equation}
where $ \varrho_{\mathbf{h} \mathbf{h}} (\lambda ) $ are just
precisely the occupation probabilities in each invariant subspace. For
the case a symmetry algebra su(2), generated by $\{ \op{S}_{x}, \op{S}_{y},
\op{S}_{z} \}$, the archetypal example of this situation is
\begin{equation}
  \mathcal{L} ( \op{\varrho} ) = 
  \op{S}_{z} \, \op{\varrho} \, \op{S}_{z} -
  \frac{1}{2}
  ( \op{S}_{z}^{2} \, \op{\varrho} + \op{\varrho} \, \op{S}_{z}^{2} ) =
  \op{S}_{z} \, \op{\varrho} \, \op{S}_{z} - \op{\varrho} \, , 
\end{equation}
which is a standard way of accounting for process that lead to a loss
of coherence without changing the level populations~\cite{Briegel:1993xy}.

Next we consider a generalized amplitude-damping channel, for which
\begin{equation}
  \op{L}_{r} = \op{e}_{- \alpha_r} \, .
\end{equation}
In the asymptotic limit, this leads to the ground state in each
invariant subspace
\begin{equation}
  \lim_{n \rightarrow \infty} \mathcal{E}^{n} ( \op{\varrho} ) =
  \sum_{\lambda} |\mathbf{h}_{\mathrm{min}}; \lambda \rangle 
  \langle \mathbf{h}_{\mathrm{min}}; \lambda | \, ,
\end{equation}
where $ |\mathbf{h}_{\mathrm{min}}; \lambda \rangle $ is the
lowest-weight state that is annihilated by all the lowering operators:
$\op{e}_{-\alpha} |\mathbf{h}_{\mathrm{min}}; \lambda \rangle = 0$. It
is clear that now no Lindblad (neither Kraus) map preserves the
diagonal operators. In particular, for su(2) symmetry, the textbook
example is
\begin{equation}
  \label{eq:dam}
  \mathcal{L} ( \op{\varrho} ) = 
  \op{S}_{-} \op{\varrho} \op{S}_{+} 
  - \frac{1}{2} 
  ( \op{S}_{+} \op{S}_{-} \op{\varrho} + 
  \op{\varrho} \op{S}_{+} \op{S}_{-} ) \, .
\end{equation}
which is a standard model for damping~\cite{Agarwal:1974rm}.

While these generalizations are more or less obvious, the
corresponding one for a depolarization channel is far from trivial. 
We claim that  such a situation must be described by the action of 
a Lindblad  operator proportional to $\op{e}_{-\alpha_r}$ followed 
by other proportional to $\op{e}_{+\alpha_r}$.  One can check that 
only in this way we asymptotically get an unpolarized state 
\begin{equation}
  \lim_{n \rightarrow \infty} \mathcal{E}^{n} ( \op{\varrho} ) =
  \sum_{\lambda} \Tr [\op{\varrho} (\lambda) ] \, \op{\openone}_\lambda 
  = \op{\varrho}_{\mathrm{unpol}} , 
\end{equation}
where the trace operation is taken in each invariant subspace.  The
associated Lindblad operator is
\begin{equation}
  \label{eq:4}
 \fl
  \mathcal{L} ( \op{\varrho} )  =  \frac{1}{2} \sum_{r}  a_{r} 
  \left (  
    2 \op{e}_{-\alpha_r} \, \op{\varrho}  \, \op{e}_{\alpha_r} +  
    2 \op{e}_{\alpha_r} \, \op{\varrho}  \op{e}_{-\alpha_r} - 
    \{ \op{e}_{\alpha_r},  \op{e}_{- \alpha_r} \} \, \op{\varrho} -  
    \op{\varrho} \, \{ \op{e}_{\alpha_r},  \op{e}_{- \alpha_r} \} 
  \right ) \, ,
\end{equation}
and one can check that $\mathcal{L} ( \op{\varrho}_{\mathrm{unpol}} ) = 0$.
An example of this situation for qubit systems parallels completely
(\ref{eq:4}), but with the corresponding spin-like operators is
\begin{equation}
  \mathcal{L} ( \op{\varrho} ) =  
  \op{S}_{-} \op{\varrho}  \op{S}_{+} +  
  \op{S}_{+} \op{\varrho}  \op{S}_{-} - 
  \frac{1}{2}
  \left ( \{ \op{S}_{+},  \op{S}_{-} \} \op{\varrho} + 
  \op{\varrho }\{ \op{S}_{+},  \op{e}_{-} \}  \right ) \, . 
\end{equation}
It is worth noting that such a Linblad operator appears as a limit
case of decaying into a bath at infinite temperature.

An effective depolarization channel is actually a common situation 
in driven dissipative systems. Consider a typical evolution equation
\begin{equation}
  \label{eq:5}
  \dot{\op{\varrho}} = - i g [ \op{S}_{x}, \op{\varrho} ] + 
  \gamma [ \op{S}_{-} \op{\varrho} \op{S}_{+} - 
  \frac{1}{2} \left ( 
    \{ \op{S}_{+}, \op{S}_{-} \} \, \op{\varrho} + 
    \op{\varrho} \, \{ \op{S}_{+}, S_{-} \} \right )  \, ,
\end{equation}
which describes the decay (with rate $\gamma$) of an externally-driven
(with coupling constant $\kappa$) collective spin into a
zero-temperature bath~\cite{Agarwal:1974rm}. In the strong-pumping
limit ($g \gg \gamma$), the above equation can be easily
diagonalized~\cite{Klimov:2000la}: it is enough to apply the rotation
$\op{U} = \exp(i \pi \op{S}_{y} / 2)$ and to make the rotating wave
approximation. The final result is
\begin{eqnarray}
  \dot{\op{\varrho}}_{d} & = & - i g [ \op{S}_{z}, \op{\varrho}_{d} ] 
  +  \frac{\gamma}{2} 
  \left ( 2 \op{S}_{z} \op{\varrho} \op{S}_{z} -
  \op{S}_{z}^{2} \op{\varrho} - \op{\varrho} \op{S}_{z}^{2}  \right ) 
\nonumber \\
& + & \frac{\gamma }{2} \left  (
  2 \op{S}_{-} \op{\varrho} \op{S}_{+} + 
  2 \op{S}_{+} \op{\varrho} \op{S}_{-} -
  \{ \op{S}_{+}, \op{S}_{-}\} \, \op{\varrho} - 
  \op{\varrho} \,  \{ \op{S}_{+}, \op{S}_{-}\} \right ) \, ,
\end{eqnarray}
where $\op{\varrho}_{d} = \op{U} \, \op{\varrho} \, \op{U}^{\dagger}$
is the density matrix in the rotated frame. We can clearly observe the
emergence a pure dephasing (first line) and a depolarizing channel
(second line).

In summary, what we expect to have accomplished in this paper is to
provide a construction of the depolarizing channel for systems with
arbitrary symmetries. This may be more than an academic curiosity for
more involved systems currently under investigation as candidates for
quantum information processing.

\ack
  
This work was supported by the Grant No 45704 of Consejo Nacional de
Ciencia y Tecnolog\'{\i}a (CONACyT) and the Spanish Research
Directorate (Grant FIS2005-06714).  A. B. K. was also supported by the
Spanish Sabbatical Program (Grant SAB2006-0064).

\newpage


\end{document}